# Investigation of strut-ramp injector in a Scramjet combustor: effect of strut geometry, fuel and jet diameter on mixing characteristics

Rahul Kumar Soni, Ashoke De[*]

*Department of Aerospace Engineering, Indian Institute of Technology Kanpur, Kanpur, India, 208016*

___

**Abstract**

The strut based injector has been found to be one of the most promising injector designs for supersonic combustor, offering enhanced mixing of fuel and air. The mixing and flow field characteristics of the straight (SS) & tapered strut (TS), with fixed ramp angle and height at freestream Mach number 2 in conjunction with fuel injection at Mach 2.3 have been investigated numerically and reported. In the present investigation, hydrogen ($H_2$) and ethylene ($C_2H_4$) are injected in oncoming supersonic flow from the back of the strut, where jet to freestream momentum ratio is maintained at 0.79 and 0.69 for $H_2$ & $C_2H_4$, respectively. The predicted wall static pressure and species mole fractions at various downstream locations are compared with the experimental data for TS case with 0.6 mm jet diameter and found to be in good agreement. Further, the effect of jet diameter and strut geometry on the near field mixing in strut ramp configuration is discussed for both the fuel. The numerical results are assessed based on various parameters for the performance evaluation of different strut ramp configurations. The SS configuration for both the injectant is found to be an optimum candidate, also it is observed that for higher jet diameter larger combustor length is required to achieve satisfactory near field mixing.

___

## 1. Introduction

The development of high speed propulsion system requires thorough understanding of the complex flow physics associated with the combustor. Over the decade, research community has found Scramjet engines to be the most efficient air-breathing propulsion system in the high speed flow regime. However, due to intrinsic difficulties associated with the combustion mechanism in a supersonic flow, there exists a need for development in the area of fuel mixing and flame holding. At higher supersonic speeds, the residence time for ingested air within the Scramjet unit is on the order of milliseconds; in turn it means that the fuel injected in the oncoming air must mix efficiently and burn to release energy within few milliseconds. Hence, the very first step towards the enhanced combustion efficiency could be attained by designing efficient mixing strategy which can offer stable operating condition (or stabilized flame front). Various researchers over the year have explored wide range of injection mechanism that includes splitter plate, normal injection, compression/expansion ramps, lobe mixer and many more detailed reviews regarding the same can be found in [1].

The earlier injection strategy involved transverse injection onto the oncoming supersonic crossflow [2-8]. Transverse injections lead to the formation of normal bow shock separating two regions of flow i.e. upstream and downstream. A recirculation zone is created at the downstream, which aids in flame holding and thereby offering higher combustion efficiency. The total pressure loss, however due to normal shock wave is significantly high and effects the scramjet cycle performance. However, various literatures [9-16] suggest that the strut based parallel injection to be more promising, as it offers the possibility of injecting fuel into the core of oncoming supersonic flow leading to uniform spreading of the fuel. The parallel injection system is also known to offer improved cycle performance, whereas the combustion efficiency is reduced due to deterioration in near field mixing. Diamotakis [17] reported that the mixing in parallel injection system could be improvised with the generation of axial vorticity. This led to the research into the various strut designs. The presence of strut leads to the bifurcation of oncoming supersonic flow which enhances the mixing due to formation of the shear layer behind the strut. Apart from this, another important phenomenon contributing towards the mixing enhancement is the shock-shear layer interaction. The shock generated at the ramp (wedge) leading edge undergoes reflection and continuously interacts with the supersonic shear layer leading to the generation of vorticity through the baroclinic torque mechanism [18-22]. Menon & Genin [23] in their study with injection through wedge observed that mixing in non-reacting cases is mainly due to unsteadiness and enhanced level of turbulence in the shear layer. This interaction perturbs the shear layer, result of which is larger entrainment of surrounding air which also contributes towards the enhanced mixing.

Research on strut mixing devices covers a wide range of designs and includes both normal and parallel injection methodologies. Most struts consist of a ramp followed by the strut and fuel is injected from the strut. Dessornes and Jourdren [15] compared three mixing techniques for scramjet combustion: transverse injection in a cavity, two-stage transverse injection and a strut consisting of a vertical wedge front with fuel injection at the back-side of the trailing edge. They found that a strut was the only technique that affected the entire flow field due to deeper penetration but had a higher pressure loss than the other techniques. The researchers suggest that more interest should be paid to the design of the strut to minimize the pressure loss while maintaining the ability to affect the flow field.

From the few existing literatures, it has become evident that the parallel injection into the core of oncoming supersonic flow is promising. However, there is still a need to perform extensive study to understand the flow physics and mixing characteristics for several possible configurations. Various parameters need to be investigated like ramp angle, strut length and lip thickness which directly not only affects the mixing but also the length of the combustor required to achieve satisfactory mixing. To the best of author's knowledge in open literature, this type of investigation has not been reported especially involving this particular configuration; while most of the studies in the past have focused mainly on lobed struts or cantilevered type injector. Only study involving lip thickness variation as parameter for tapered strut with fixed diameter has been reported by Lee [10]. This paves the way for current investigation with multiple parameters with emphasis on strut geometry as well. Hence, the primary motivation of the present work is to characterize the mean flow features to understand the mixing behaviour. Therefore, the present study is performed with two equation RANS (Reynolds Averaged Navier Stokes) based model, as these models require lesser computational resource (cost effective) and offer good understanding of the mean flow field. In the present work, flow field and mixing characteristics for SS & TS injectors are investigated for hydrogen and ethylene. The effect of jet diameter (0.6, 1 and 2 mm) on flow features and mixing is studied for both the fuel including different strut configurations. In the case of SS configuration, only lip thickness varies due to jet diameter variation; however in case of TS, both the taper angle and lip thickness vary to accommodate the given jet diameter. Initially, the validation and grid independence of the solver is demonstrated followed by the detailed parametric investigation on mixing characteristics.

## 2. Numerical Details

### 2.1 Governing Equations

The Favre-averaged governing equations for fluid motion are discretized and solved using the Finite-Volume method (FVM). The equations for continuity, momentum, energy, species transport and turbulence transport are recast as:

Continuity:

$$\frac{\partial \bar{\rho}}{\partial t} + \frac{\partial \bar{\rho}\tilde{u}_i}{\partial x_i} = 0 \quad (1)$$

Momentum:

$$\frac{\partial}{\partial t}(\bar{\rho}\tilde{u}_i) + \frac{\partial}{\partial x_j}[\bar{\rho}\tilde{u}_i\tilde{u}_j + \bar{p}\delta_{ij} + \overline{\rho u_i'' u_j''} - \bar{\tau}_i] = 0 \quad (2)$$

Energy:

$$\frac{\partial}{\partial t}(\bar{\rho}\tilde{E}) + \frac{\partial}{\partial x_j}[\bar{\rho}\tilde{u}_j\tilde{E} + \tilde{u}_j\bar{p} + \overline{u_j'' p} + \overline{\rho u_i'' E''}$$
$$+ \overline{q_j} - \overline{u_i \tau_{ij}}] = 0 \quad (3)$$

Species Transport:

$$\frac{\partial}{\partial t}(\bar{\rho}\tilde{Y}_k) + \frac{\partial}{\partial x_j}(\bar{\rho}\tilde{u}_j\tilde{Y}_k) -$$
$$\frac{\partial}{\partial x_j}\left(\overline{\rho D_k \frac{\partial Y_k}{\partial x_j}} - \overline{\rho u_i'' Y_k''}\right) = 0 \quad (4)$$

Where (~) & (-) refers to density weighted time averaging and averaging through Reynolds decomposition respectively. $Y_k$ is the mass fraction of kth specie $\tau_{ij}$, E and $S_{ij}$ are shear stress, strain rate and total energy calculated as,

$$\overline{\tau_{ij}} = 2\mu S_{ij} - \frac{2}{3} S_{kk} \delta_{ij} \tag{5}$$

$$S_{ij} = \frac{1}{2}\left(\frac{\partial u_i}{\partial x_j} + \frac{\partial u_j}{\partial x_i}\right) \tag{6}$$

$$\tilde{E} = e + \frac{\tilde{u}_k \tilde{u}_k}{2} + k \tag{7}$$

$$-\overline{\rho u_i'' u_j''} = \mu_t \left(\frac{\partial u_i}{\partial x_j} + \frac{\partial u_j}{\partial x_i}\right) + \frac{2}{3}\rho k \delta_{ij} \tag{8}$$

The $\mu_t$ in eq. (8) is evaluated by complementing the above set of equation along with the transport equations of turbulence quantities. In the present study, SST k-ω model proposed by Menter (1994) is utilized to close the above set of equations. The eddy viscosity μt is given as,

$$\mu_t = \frac{\rho a_1 k}{max[a_1 \omega, \Omega F_2]} \quad , \quad a_1 = 0.31 \tag{9}$$

$$F_2 = tanh\left\{\left(max\left[2\frac{\sqrt{k}}{0.09\omega y}, \frac{500\mu}{\rho y^2 \omega}\right]\right)^2\right\} \tag{10}$$

$$\frac{\partial(\rho k)}{\partial t} + \frac{\partial}{\partial x_j}\left(\rho u_j k - (\mu + \sigma_k \mu_t)\frac{\partial k}{\partial x_j}\right) = \tau_{ij}\frac{\partial u_i}{\partial x_j} - \beta^* \rho \omega k \tag{11}$$

$$\frac{\partial(\rho \omega)}{\partial t} + \frac{\partial}{\partial x_j}\left(\rho u_j \omega - (\mu + \sigma_\omega \mu_t)\frac{\partial \omega}{\partial x_j}\right) = P_\omega - \beta \rho \omega^2$$

$$+2(1-F_1)\frac{\rho\sigma_{\omega 2}}{\omega}\frac{\partial k}{\partial x_j}\frac{\partial \omega}{\partial x_j} \quad (12)$$

$$P_\omega = \frac{\gamma}{\nu_t}\tau_{ij}\frac{\partial u_i}{\partial x_j} \quad (13)$$

$$\tau_{ij} = \mu_t\left(2S_{ij} - \frac{2}{3}\frac{\partial u_k}{\partial x_k}\delta_{ij}\right) - \frac{2}{3}\rho k \delta_{ij} \quad (14)$$

$$\sigma_k = 0.85, \quad \sigma_{\omega 2} = 0.856$$

$$F_1 = tanh\left\{\left(min\left[max\left\{\frac{\sqrt{k}}{0.09\omega y}, \frac{500\mu}{\rho y^2 \omega}\right\}, \frac{4\rho\sigma_{\omega 2} k}{CD_{k\omega} y^2}\right]\right)^4\right\} \quad (15)$$

$$CD_{k\omega} = max\left[\frac{2\rho\sigma_{\omega 2}}{\omega}\frac{\partial k}{\partial x_j}\frac{\partial \omega}{\partial x_j}, 10^{-20}\right] \quad (16)$$

$\Omega$ in eq. (9) $\Omega$ is the vorticity magnitude and F2 is a blending function given by eq. (10). The last term in the eq. (12) is cross diffusion term evaluated through eq. (16) and production of ω is approximated through eq. (13). Similar to $F_2$, $F_1$ is also a blending function given by eq. (15). Further details about the Menter's model can be found in literatures [24-26].

*2.2 Numerical Scheme*

In the present work, the density based solver in OpenFOAM framework is modified to accommodate the transport of multi-species system, which is based on finite volume discretization utilizing semi-discrete, non-staggered central schemes for co-located variables on polyhedral mesh. The transport equations are solved using operator-splitting approach where initially, convection of conserved variables are solved through explicit predictor equation and then diffusion of primitive variables are solved using implicit corrector equation. The solver utilizes central schemes proposed by [20 & 21] which is an alternative approach to Riemann solver offering accurate non-oscillatory solution. More detailed information about the implementation in OpenFOAM can be found in [14]. In present simulation, second order backward Euler scheme is utilized for the time integration whereas viscid and inviscid fluxes are discretised using central difference and TVD scheme. The parallel processing is achieved through the message passing interface (MPI) technique.

*2.3 Boundary Details*

At inlet, fixed value is defined for all the variables as tabulated in Table 1. At the solid surface no-slip boundary condition is imposed, while at the outlet zero-gradient is used for all the variables excluding pressure for which a non-reflecting boundary condition is imposed to avoid the incoming waves to enter the domain. One should note that the substantial difference in inlet velocity of hydrogen and ethylene (for same Mach number) is primarily due to the difference in specific heat ratio and gas constant. The Lewis number is assumed to be 1 in the present study. The simulation is carried out for 30 non-dimensional times while maintaining CFL number below 0.5.

The density based solver is modified to accommodate species transport equation and then validated against the experimental results of Gerlinger and Brüggemann (2000). The TS-0.6 mm configuration with $H_2$ as injectant is presented for solver validation and then detailed results for all the cases are discussed. The jet to freestream momentum ratio is 0.79 and 0.69, for the $H_2$ and $C_2H_4$ respectively. The ratio is defined as Jet momentum flux ratio = $(\gamma PM^2)_j/(\gamma PM^2)_\infty$.

Table 1. Inlet Conditions for main and jet flow

| Parameter | Air | Ethylene | Hydrogen |
|---|---|---|---|
| $P_\infty$, Pa | 49.5 | 29.5 | 29.5 |
| $T_\infty$, K | 159 | 151 | 151 |
| $M_\infty$ | 2 | 2.3 | 2.3 |
| $U_\infty$, m/s | 505 | 556 | 2203 |
| $Y_{N_2}$ | 0.76699 | 0 | 0 |
| $Y_{O_2}$ | 0.23301 | 0 | 0 |
| $Y_{C_2H_4}$ | 0 | 1 | 0 |
| $Y_{H_2}$ | 0 | 0 | 1 |

Table 2. Grid spacing details

| | Jet Region (mm) | Wall Region (mm) |
|---|---|---|
| Grid 1 | $\Delta x = 8\times10^{-03}$ | $\Delta x = 5\times10^{-03}$ |
| | $\Delta y = 6\times10^{-04}$ | $\Delta y = 8\times10^{-04}$ |
| Grid 2 | $\Delta x = 5.1\times10^{-04}$ | $\Delta x = 5.1\times10^{-04}$ |
| | $\Delta y = 3\times10^{-05}$ | $\Delta y = 1.4\times10^{-04}$ |
| Grid 3 | $\Delta x = 2.8\times10^{-05}$ | $\Delta x = 2.8\times10^{-05}$ |
| | $\Delta y = 1.4\times10^{-05}$ | $\Delta y = 8.17\times10^{-05}$ |

## 3. Results & Discussion

*3.1 Grid Independence and Study*

The grid is generated by creating multiple blocks to accommodate varying grid size using stretching function to maintain optimum grid size while resolving the important flow feature such as recirculation zone, shock, shock-shock interaction and shock-shear interaction. To perform the grid independence study, three sets of grids are generated namely Grid1, Grid2 and Grid3 with 650×110, 1200×240 and 1950×450 cells, respectively.

The current flow involves shock train throughout the channel length, which introduces sharp gradient along both streamwise and transverse direction. The grid for present computation is designed in such a manner that the near wall region is sufficiently resolved and similarly the near jet region, in between the two regions uniform stretching is utilized to keep the grid size optimum. The Grid 1 has larger mesh gradient along the transverse direction and this behavior can be attributed to the grid resolution along the transverse direction. Furthermore, the details of grid spacing are provided in Table 2 for better understanding.

The normalized velocity distribution for TS-0.6 mm case with hydrogen injection is presented in Figure 2 to demonstrate the influence of grid on numerical simulation. The experimental results are not available for the velocity; hence the quantitative comparison for the velocity field cannot be made. However, we have presented the quantitative comparison for $H_2$ mole fraction along with the velocity field for better assessment (Fig. 2).

It is quite evident from this figure that the results using both the Grid 2 & Grid 3 offer good agreement with the experimental data which can be verified from the species mole fraction, and the discrepancies in Grid 1 is primarily due to the improper predictions of velocity field.

The velocity distribution for Grid 2 & 3 follows closely but Grid1 shows under prediction especially at the downstream locations, in turn this grid (Grid1) under predicts the entrainment of the surrounding fluid. Also, qualitatively Grid1 under predicts the recirculation zone present on the either side of the jet exit. The presence of the re-circulation region compresses the jet which then tends to spread out in cross-streamwise direction, downstream immediately. The predictions using Grid2 & Grid3 are in excellent agreement with each other; hence Grid2 is chosen for rest of the detailed computations and discussed hereafter.

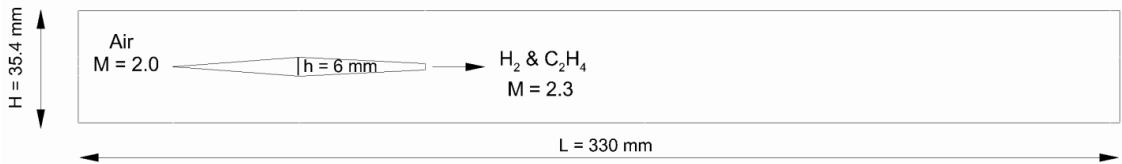

Fig. 1. Computation domain with dimension for TS -0.6 case

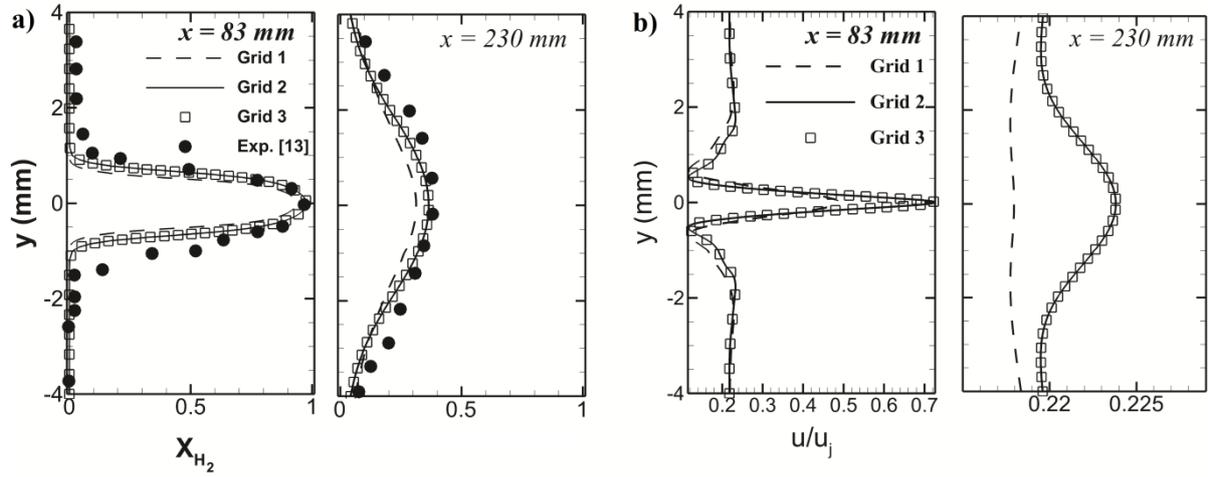

Fig. 2. a) Mole Fraction & b) velocity profile for grid independence demonstration at two axial locations

The wall pressure distribution presented in Figure 3 is compared to validate the solver alongside the observation of [13]. It is evident that both the numerical results follow closely and are in good agreement with the experimental data; however, there exists subtle differences as the former one [13] used different RANS models. Also, species mole fraction profiles of hydrogen and nitrogen are compared at four different locations which correspond to x = 83, 130 and 230 mm as depicted in Figure 4.

As observed, at x = 83 mm, both nitrogen and hydrogen follow the experimental trend very nicely; however as we move further downstream the jet diffusion in y-direction appears to be slightly under predicted but still yields reasonably accurate results. The difference along the wall normal direction can be attributed to the shift in shock locations which is consistent with the observation of [13] and builds the scope for future investigation with high end turbulence models.

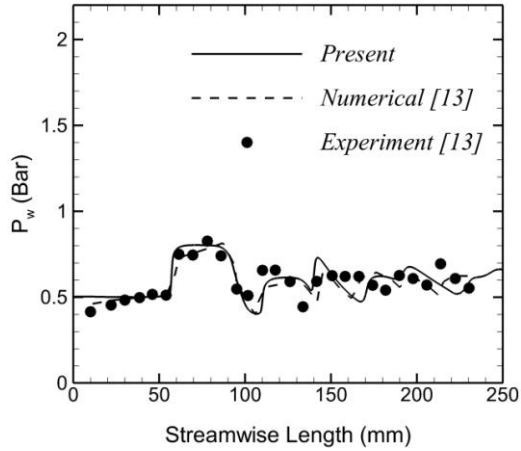

Fig. 3. Wall pressure distribution compared with numerical and experimental observation of [13]

*3.2 Flow Features*

The density gradient contour is presented in Figure 5, where the complex flow features can be observed. Oblique shock wave is generated as the oncoming supersonic flow encounters the ramp which is then reflected from the walls. Some of the main flow features observed are shock/shock, shock/boundary layer and shock/shear interaction. The reflected shock wave upon interacting with the shear layer leads to the production of the vorticity which leads to jet break up and aids in near field mixing. The flow on reaching the edge of strut reverses due to the finite lip thickness which reattaches through the reattachment shock. This reattachment shock also contributes towards the vorticity production.

Another interesting phenomenon which is evident from the Figure 5 is the interaction of shock wave and boundary layer along the channel wall. This interaction leads to the separation of boundary layer and vortical shedding past the separation point. In transverse injection where complicated shock structures (bow & lambda shock) are present the separation of boundary layer due to shock formation introduces vortex shedding downstream of the jet injection.

These vortices transport the injectant and enhance the air/fuel mixing [38]. However in the present case shock/boundary layer interaction will have little or no effect on the mixing. Hence, the resolution of SWBLI is not of much relevance in the current computation.

The reattachment length and hence the reattachment shock strength varies depending on the lip thickness. The strength of recir-culation also affects the mixing by altering the mixing layer thickness, larger recirculation zones are known to introduce higher level of turbulence. The extent of recirculation zone created at the strut corner has direct impact on the mixing and jet spreading which will be discussed in the following section.

In Figure 6, the close up views of jet for TS & SS configuration with 0.6 mm jet diameter is presented to get a qualitative estimate of the reattachment length. Evidently, huge difference can be observed in the size of the recirculation zone. For TS-0.6 mm case, the reattachment length is approximately 1.8 mm but for SS-0.6 mm case much larger recirculation zone is observed, approximately 4.6 mm. At the point of reattachment, the jet core appears to be compressed due to the presence of the reattachment shock. This particular difference has lot of impact in the downstream flow features as discussed in the following sub-sections.

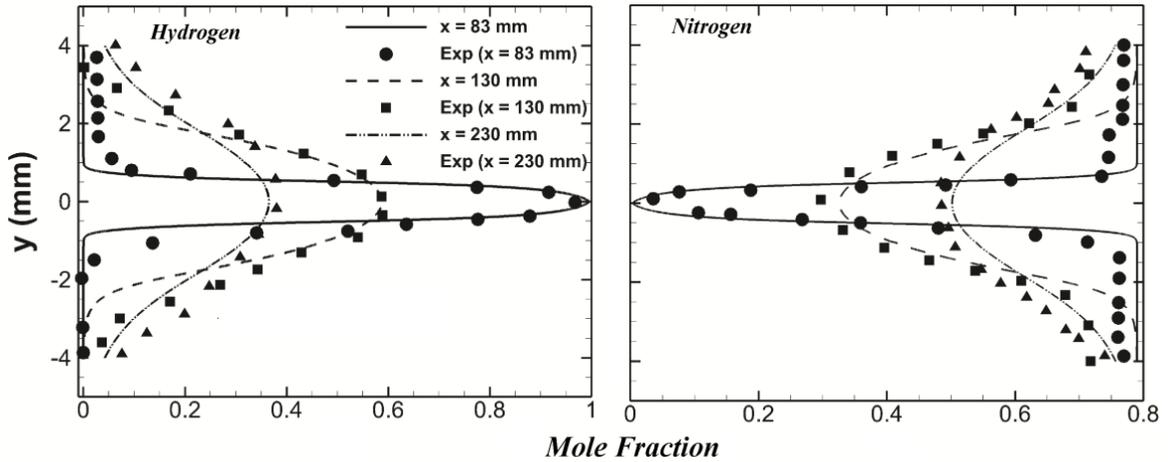

Fig. 4. Comparison of $H_2$ and $N_2$ mole fraction profile at various streamwise locations downstream of jet

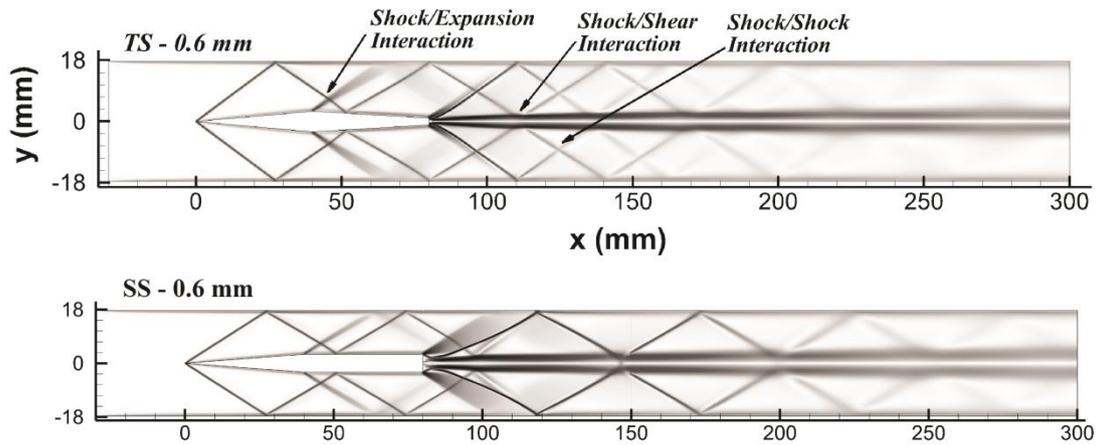

Fig. 5. Numerical Schlieren ($\nabla \rho$) showing complex flow physics

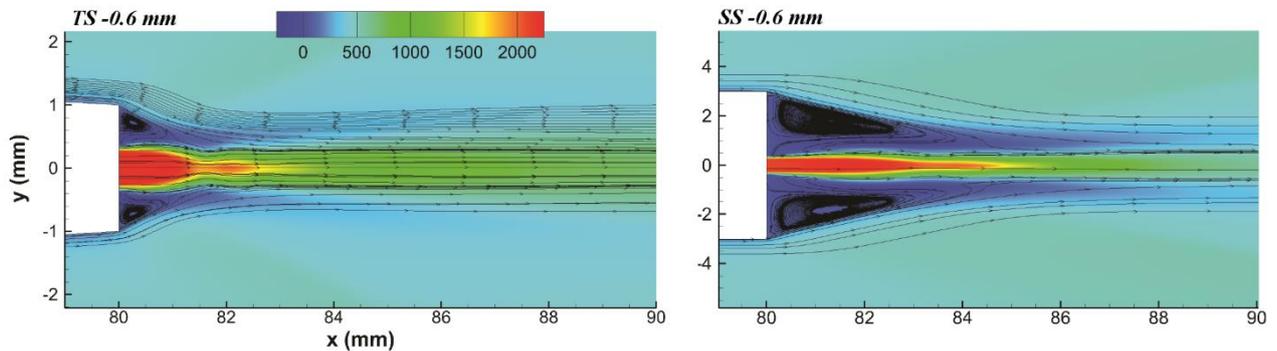

Fig. 6. Close up view of 0.6 mm jet for both SS & TS configuration

*3.3 Effect of Fuel & Geometry*

For detailed parametric study, computations are performed for two different strut geometries including three different jet diameters using two different fuels ($H_2$ & $C_2H_4$). It is to be noted that for all the cases reported in this section, the length and wedge angle are maintained constant. Figure 7 shows the predicted hydrogen mole fraction at different stream-wise locations; where for a given jet diameter, the results of TS & SS cases is plotted for the comparison. With the increasing jet diameter the penetration in cross stream-wise direction enhances, i.e. jet spreading is more in radial direction; while the significant difference can be witnessed for the 0.6 mm and 1 mm cases for both TS & SS. However with increasing jet diameter the performance along the stream-wise direction appears to deteriorate with the larger presence of hydrogen mole fraction. In case of 0.6 mm diameter, difference between TS & SS configuration is more significant than 1 & 2 mm cases. The SS-0.6 mm case appears to have better performance in terms of near field mixing compared to TS-0.6 mm due to the significant difference in the extent of the recirculation region. In case of 1 & 2 mm cases for both the strut configurations, the difference is not very significant due to the presence of comparable recirculation zone. Overall, the stream-wise distribution for higher diameters for both the cases suggests that for higher jet diameter larger combustor length might be required to allow the proper mixing. But this can be remedied by producing stronger shock and hence the baroclinic torque which will aid in early jet breakup and hence mixing augmentation.

Similarly, Figure 8 presents the ethylene mole fraction distribution at four different stream-wise locations for both the strut geometries. In contrast to Figure 7, the ethylene seems to have better performance along the stream-wise direction but lesser diffusion in transverse direction.

The reason behind poor jet spreading in cross-stream direction could be attributed to the smaller velocity gradient between primary and secondary jet which directly effects the entrainment from secondary flow. However, in the case of hydrogen injection the velocity gradient along the trans- verse direction is higher as the velocity of primary flow for same Mach number is approximately four times that of ethylene.

The velocity profile along the downstream locations is another parameter which throws light on the behavior of jet evolution and mixing characteristics. In Figure 9, the normalized velocity profiles is presented, the normalization for both the cases are done by the fuel jet exit velocity of the respective fuels, details of which are provided in Table 1.

At *x = 83 mm* it can be seen that for all the cases the symmetric profile initially remains constant ($\pm 2 < y < \pm 4$) and upon reaching the recirculation region ($0 < y < \pm 2$) it reduces and then increase towards the core of jet to attain the maximum velocity. Worth noticing is the larger and wider low velocity region for the straight strut cases (SS) compared to the tapered strut cases (TS). When comparing TS & SS cases particularly at *x = 83 mm*, it can be noticed that the flow accelerates just before it reaches the recirculation region, as observed for both the injection cases, but this acceleration is more pronounced for the ethylene cases. Furthermore, it can be observed that with the increasing jet diameter this acceleration of the flow is decreased outside the recirculation region on either side of jet; which is primarily due to the reduction in the taper angle for the higher jet diameter as the flow turns through the expansion fan, strength of which decreases with the taper angle. At other locations further downstream of the jet injection *x = 83 mm*, for $H_2$ injection the velocity profile tends to become more flat due to the larger entrainment from the surrounding secondary flow compared to the $C_2H_4$ and this can be noticed for both the strut geometries.

*3.4 Total Pressure Decay*

In any study involving supersonic mixing or combustion, two parameters widely accepted for the characterization of any injection technique are total pressure & temperature decay suggested by [7, 15 & 36]. However due to freestream temperature of primary and secondary flow being approximately same, total temperature decay cannot be a good indicator of mixing characteristics. Total temperature decay is primarily a good parameter for studies involving significant temperature difference. Hence in the present case only total pressure decay along the jet centerline is discussed.

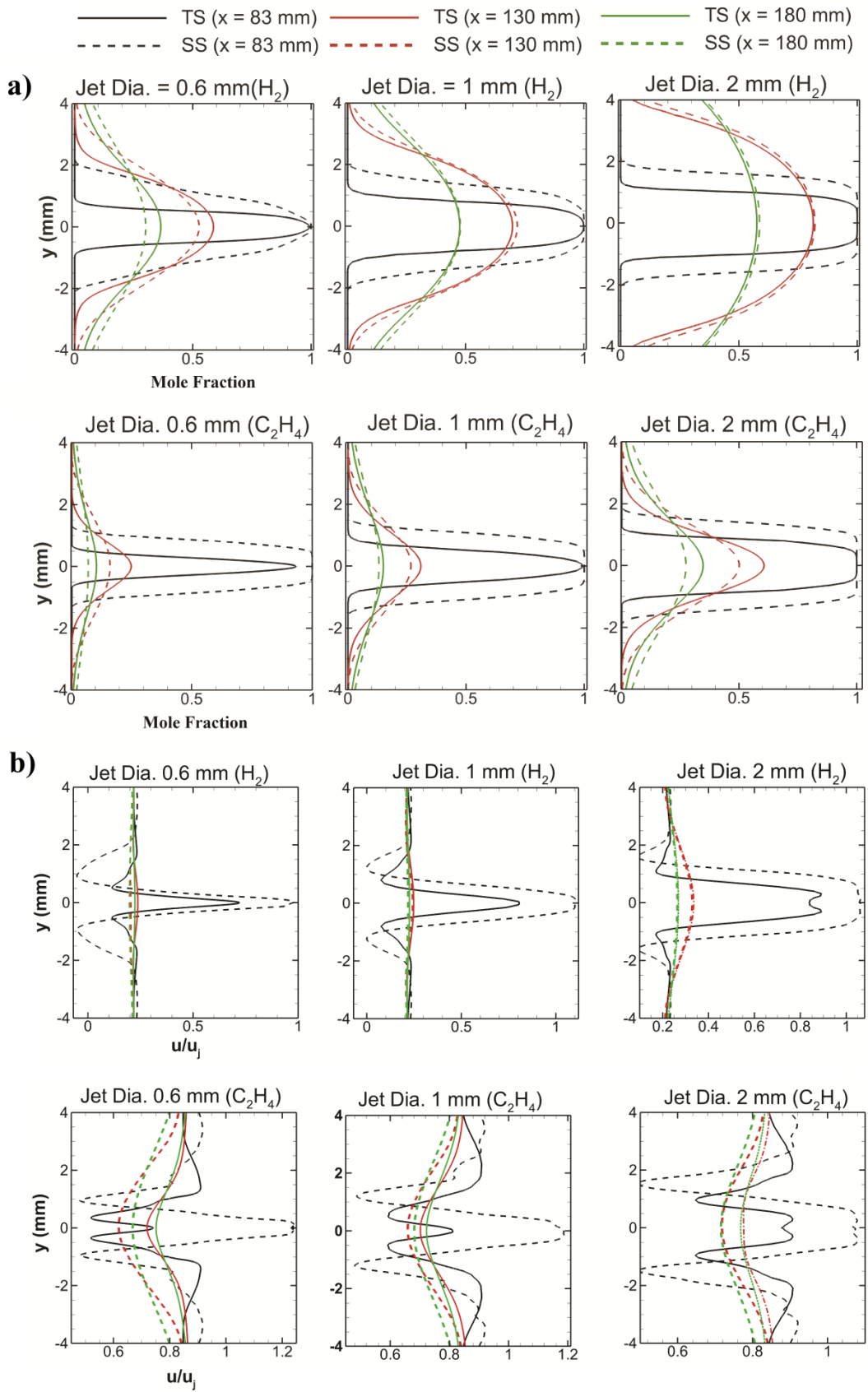

Figure 8: a) Mole Fraction and b) Velocity profile at different locations

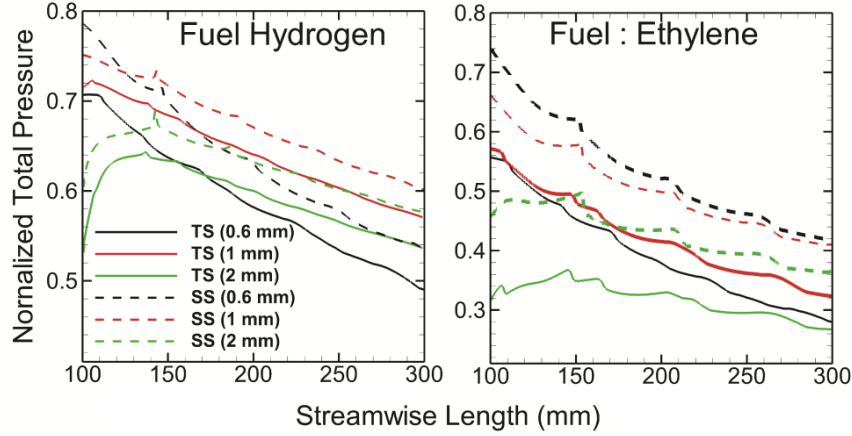

Fig. 10. Total pressure decay for both $H_2$ & $C_2H_4$ for varying jet diameter for both the strut

Total pressure decay (TPD) is a good indication of the spreading rate and mixing characteristics. The increased decay in total pressure distribution is indication of higher spreading rate and enhanced mixing rate. Figure 10 presents normalized total pressure along the jet centerline where stagnation pressure is normalized as $(P_{o,j} - P_o)/P_{o,j}$.

It can be seen that the decay is higher for the 0.6 mm case with $H_2$ injection, however worth noticing is the difference between TS-0.6 and SS-0.6. This difference suggests that the SS-0.6 case has slightly higher spreading rate which can be confirmed from the Figure 6. As observed in Figure 6, the growth rate of jet is slightly more compared to TS-0.6. This is due to the fact that the stronger reattachment shock is created in case of SS-0.6 mm case which can be verified from Figure 6, compared to other two which is consistent with the observation of [13].

This observation is consistent with the observations of velocity and mole fraction profiles. Overall, hydrogen appears to have the desirable mixing and spreading rate for the supersonic combustion. Also it can be inferred that hydrogen would require shorter length compared to ethylene for better mixing. The increased entrainment of secondary flow in the primary flow is higher for hydrogen for all jet diameters while the same appears to be very less for ethylene case. It is noteworthy to mention that the present investigation reveals that overall SS configuration seems to perform better than the TS cases for all the jet diameters as it offers stronger recirculation region in the vicinity of jet exit.

Keeping the above mentioned discussion in mind, the most promising configuration from the present study appears to be SS-0.6 mm cases. From the total pressure decay observation it is found that for 0.6 mm $H_2$ case, SS configuration has around 11% more total pressure decay than the TS-0.6 mm configuration which is evident form the species molar fraction distribution. Similarly in case of $C_2H_4$ injection SS-0.6 mm offers 19% more total pressure decay than the TS-0.6 mm case. For higher jet diameter the percentage change is almost negligible, especially in case of 2 mm injection. In general, SS configuration appears to be more promising however with increasing diameter combustor length required to achieve near field mixing appears to be increased; while for higher jet diameter the SS configuration might require stronger shock to keep the combustor length optimum.

### 3.5 Mixing Efficiency

Finally mixing efficiency ($\eta_m$) for all geometrical variation and both fuel is computed based on the relation coined at NASA Langley Research Center [37]. The two part definition is given in equation 17 & 18.

$$\eta_m = \frac{\int Y_f \rho u \, dA}{\int Y \rho u \, dA} \qquad (17)$$

where,

$$Y_f = \begin{cases} Y, & Y \leq Y_s \\ Y_s(1-Y)/(1-Y_s), & Y > Y_s \end{cases} \qquad (18)$$

Mixing efficiency is that fraction of least available reactant that can react is the mixture is brought to chemical equilibrium. The value of $\eta_m$ varies from 0 to 1, where $\eta_m = 0$ represents perfectly segregated jet whereas $\eta_m = 1$ is the indication of perfectly mixed system.

Figure 11 presents the mixing efficiency for both $H_2$ & $C_2H_4$ for all jet diameters corresponding to both geometries. Simi-lar to earlier observations here once again it can be noticed that SS-0.6 mm for both ethylene and hydrogen exhibits excellent mixing characteristics. It can be seen that for 0.6 mm case SS requires lesser combustor length compared to TS case. It can be seen that SS-0.6 ($H_2$) achieves 100% mixing efficiency at about 225 mm which is also same for SS-0.6 ($C_2H_4$); however TS-0.6 ($H_2$) achieves similar efficiency at around 275 mm, whereas TS-0.6 ($C_2H_4$) requires 300 mm to attain similar mixing efficiency. For higher diameters similar observation is noticed, in case of hydrogen injection the mixing efficiency for 1 mm case is almost 60% whereas it decreases to 40% for the 2 mm case. This observation again suggests the need for longer combustor to achieve near field mixing or stronger shock generation through ramp for 1 & 2 mm diameter.

At last, the quantitative comparison for all cases is presented in table 3. For comparing the performance of different combination mixing efficiency, combustor length and total pressure decay (TPD) is presented. The TPD here is the difference between jet exit and domain outlet, calculated from figure 10. From table 3 it is evident that over all SS configuration performs better than the TS configuration for both the fuels. However the increase of jet diameter can be witnessed from the combustor length required for complete mixing.

The total pressure decay from table 3 points out that for the entire cases hydrogen injection offers better spreading rate compared to the ethylene, as for all cases the TPD is always higher as opposed to the ethylene injection. The increased decay points toward the enhanced entrainment of primary air in the jet region which is consistent with the observation of Figure 9.

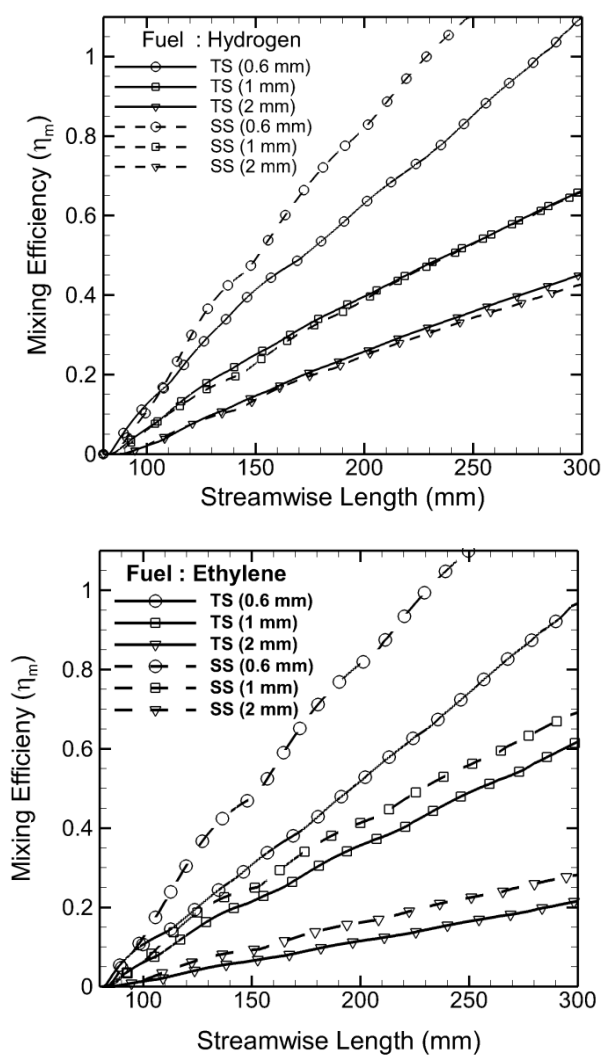

Fig. 11. Mixing Efficiency along the jet centerline

Table3: Quantitative comparison of various parameters

| Fuel | Dia.(mm) | $\eta_m$ (%) | Length (mm) | TPD |
|---|---|---|---|---|
| **H₂** | **TS** | | | |
| | 0.6 | 100 | 275 | 0.27 |
| | 1 | 60 | 300 | 0.21 |
| | 2 | 40 | 300 | 0.02 |
| | **SS** | | | |
| | 0.6 | 100 | 225 | 0.30 |
| | 1 | 60 | 300 | 0.25 |
| | 2 | 40 | 300 | 0.06 |
| **C₂H₄** | **TS** | | | |
| | 0.6 | 100 | 225 | 0.21 |
| | 1 | 60 | 300 | 0.12 |
| | 2 | 20 | 300 | 0.01 |
| | **SS** | | | |
| | 0.6 | 97.5 | 300 | 0.25 |
| | 1 | 64 | 290 | 0.15 |
| | 2 | 25 | 300 | 0.02 |

## 4. Conclusions

Numerical simulation of supersonic planar jet has been performed for two configurations with hydrogen and ethylene as fuel. The validation and grid independence study is performed and the acceptable accuracy in the numerical realization has given confidence for further detailed study. Also, the effects of strut geometry for different planar jet with two fuels are performed. The observation of overall exercise is provided categorically:

  i. Wall pressure and radial profile of mole fractions are found to be satisfactory. The difference in reattachment region due to lip thickness appears to play crucial role in mixing phenomenon.
 ii. The mole fraction profiles for both fuel suggest hydrogen to be better candidate for given strut geometry. SS configuration with hydrogen is found to be more promising. Better spreading for fuel jet is observed for SS configuration but for higher jet diameter the performance deteriorates especially in case of 2 mm. In general, it can be inferred that mixing effectiveness decreases with increasing jet diameter for a given configuration.
iii. Total pressure decay is higher for hydrogen case, suggesting better mixing efficiency and spreading rate, especially for the SS cases.
 iv. Mixing efficiency results are consistent with the total pressure decay observations.
  v. Taper angle and extent of recirculation region has direct effect on the near-field mixing.

**Acknowledgment**

Financial support for this research is provided through IITK-Space Technology Cell (STC). Also, the authors would like to acknowledge the High Performance Computing (HPC) Facility at IIT Kanpur (www.iitk.ac.in/cc).

## Nomenclature

| | |
|---|---|
| $\rho$ | : Density |
| T | : Temperature |
| P | : Pressure |
| $u_i, u_j$ | : Velocity |
| $\delta_{ij}$ | : Kronecker delta |
| $\mu_t$ | : Eddy Viscosity |
| $\omega$ | : specific dissipation rate |
| k | : Turbulent Kinetic Energy |
| $\gamma$ | : specific heat ratio |
| M | : Mach number |
| $\eta_m$ | : Mixing Efficiency |
| $\chi$ | : species mole fraction |
| $Y_f$ | : species mass fraction |
| $Y_s$ | : stoichiometric mass fraction |
| | 0.0292 for $H_2$ - air |
| | 0.068 for $C_2H_4$ - air |

**Subscripts**

| | |
|---|---|
| o | : Stagnation Condition |
| $\infty$ | : Freestream condition |
| j | : Jet exit condition |